# Spin-phonon coupling in BaFe$_{12}$O$_{19}$ M-type hexaferrite

Flávio M. Silva Júnior[1] and Carlos W. A. Paschoal[*,1]

Departamento de Física, Universidade Federal do Maranhão, Campus do Bacanga,

65085-580, São Luis-MA, Brazil

## Abstract

The spin-phonon coupling in magnetic materials is due to the modulation of the exchange integral by lattice vibrations. BaFe$_{12}$O$_{19}$ M-type hexaferrite, which is the most used magnetic material as permanent magnet, transforms into ferromagnet at high temperatures, but no spin-phonon coupling was previously observed at this transition. In this letter, we investigated the temperature-dependent Raman spectra of polycrystalline BaFe$_{12}$O$_{19}$ M-type hexaferrite from room temperature up to 780 K to probe spin-phonon coupling at the ferrimagnetic transition. An anomaly was observed in the position of the phonon attributed to the Fe$^{(4)}$O$_6$ octahedra, evidencing the presence of a spin-phonon coupling in BaM in the ferrimagnetic transition at 720 K. The results also confirmed the spin-phonon coupling is different for each phonon even when they couple with the same spin configuration.

[*] Corresponding Author. E-mail: paschoal.william@gmail.com

M-type ferrites play a leading role in applications involving permanent magnets, covering more than 90% of the market [1]. This is mainly due to their high electrical resistivity and low magnetic losses, which enables its application in magnetic and microwave devices. One M-type hexaferrite that has been widely is the barium hexaferrite, often called BaM, that has also interesting additional properties as large magnetic anisotropy, high saturation magnetization, high chemical stability and high ferrimagnetic transition ($T_C$ = 723 K) [2]. Thus, BaM is widely applied as permament magnets, particulate media for magnetic recording, and also in multilayers chip inductors operating in GHz frequency range [3–5] [6,7].

BaM, as all M-type hexaferrites, at room temperature crystallizes in a structure which often is described in terms of two structural blocks: the spinel S block $(Fe^{3+}_2Fe^{3+}_4O_8)^{2+}$ and the R block $(BaFe_6O_{11})^{2-}$, which are stacked in a SRS*R* way to form the structure, where * indicates a 180° rotation of the block around the c-axis. This structure is known as magnetopumblite, whose symmetry belongs to the space group $P6_3/mmc$ with 64 atoms in unit cell. In this structure the $Fe^{3+}$ ions, which have spin magnetic moment 5/2, are distributed within three kinds of octahedral sites, one tetrahedral site, and one bipyramidal site. This five magnetically non-equivalent site occupation by the iron ions leads to a complex magnetic structure that exhibits a ferimagnetic phase up to 723 K, when it transforms into a paramagnetic one [8]. Also, it was recently reported multiferroicity in BaM, being observed ferroelectric hysteresis loops at room temperature [9] as well as in Sc-doped BaM, in which a conical state was stabilized up to above room temperatures tunning the Sc-dping concentration [10]

Although spin-phonon coupling is frequently investigated and observed in other multiferroic and magnetoelectric compounds [11–15], in hexaferrites there are few works reporting the observation of this phenomenon, being observed only for BaM at low temperatures [16] and recently for Z-type hexaferrites at high temperatures [17]. However, spin-phonon coupling plays an important role in hexaferrites, once it is related to magnetoelectric effect in Z-type hexaferrites [18].

In this letter we probed the spin-phonon coupling in BaM hexaferrite by Raman spectroscopy at high temperature. The Raman spectra for this structure were firstly obtained by Kreisel *et al* for BaM single crystals at room temperature [19], and more recently by Chen *et al* [16] for low temperatures. Kreisel *et al* also obtained the room-temperature Raman spectrum for BaM thin films synthesized by chemical vapor deposition (CVD) [20]. However, at the best of our knowledge, ceramics samples had not been investigated through Raman technique, mainly at high temperatures, where the Raman spectra are usually low intensity and a spin-phonon coupling is difficult to be observed.

Polycrystalline BaM was prepared by conventional solid state reaction method according to a stoichiometric mixture of $BaCO_3$ and $Fe_2O_3$ starting oxides. After mixing, the oxides were grounded using an agatha mortar and pestle, and sieved to obtain a homogeneous-sized powder. The powder was calcined at 1200 °C under air atmosphere for 4h. The purity and crystalline structure of sample were probed by X-ray powder diffraction (XRPD) using a Bruker D8 Advance with Cu-Kα radiation (40 kV, 40 mA) over a range from 20° to 100° (0.02°/step with 0.3s/step). The powder diffraction pattern was compared with data from ICSD (Inorganic Crystal Structure Database, FIZ

Kalsruhe and NIST) International diffraction database (ICSD# 201654), which confirmed the formation of pure magnetoplumbite BaM phase. For Raman spectroscopy measurements, the polycrystalline powder was mixed with a 2% solution of PVA (polyvinyl alcohol) used as binder and uniaxially pressed into 13mm pellets at 4 metric tons. The pellets were placed in alumina crucibles with sacrificial powder of same composition and sintered under air at 1200°C for 16h to obtain dense ceramics. Raman spectroscopy measurements were performed in a Raman spectrometer Horiba model iHR550 coupled to a microscope Olympus BX41 equipped with a long-working distance objective (20x,20.5 mm) lens. The 662 nm line of a diode laser operating at 17mW was used to excite the Raman signal, which was collected in an air-cooled Synapse CCD. High temperature measurements were performed by furnace TS 1200 LIKAM model in the range 300 K to 780 K.

Figure 1a shows the room-temperature Raman spectrum of BaM, in which 15 modes can be clearly observed. At room temperature group theory predicts 42 Raman-active phonons in BaM, whose distribution in terms of the irreducible representation of the $6/mmm$ factor group is $11A_{1g} \oplus 14E_{1g} \oplus 17E_{2g}$. This spectrum is similar to that observed by Kreisel *et al* [19], which performed a consistent assignment of the modes, permitting us to obtain a good comparative phonon assignment. Therefore, the band at 721 cm$^{-1}$ is a Fe$^{(3)}$O$_4$ tetrahedral stretching, while the mode at 686 cm$^{-1}$ is associated with the bipyramid FeO$_5$, both with $A_{1g}$ symmetry. The band at 618 cm$^{-1}$ is due to a octahedra mode mix of $E_{1g}$ and $E_{2g}$. The mode at 470 cm$^{-1}$ is attributed to Fe$^{(4)}$O$_6$ octahedra, with $A_{1g}$ symmetry, as well as the bands at 412 cm$^{-1}$ and 339 cm$^{-1}$, which are due to the Fe$^{(5)}$O$_6$ octahedra and mixed octahedral vibration, respectively. Finally, the bands at 186 and 174 cm$^{-1}$ are modes of the whole spinel blocks with symmetry $E_{1g}$. The classification of these phonons is summarized in Table I.

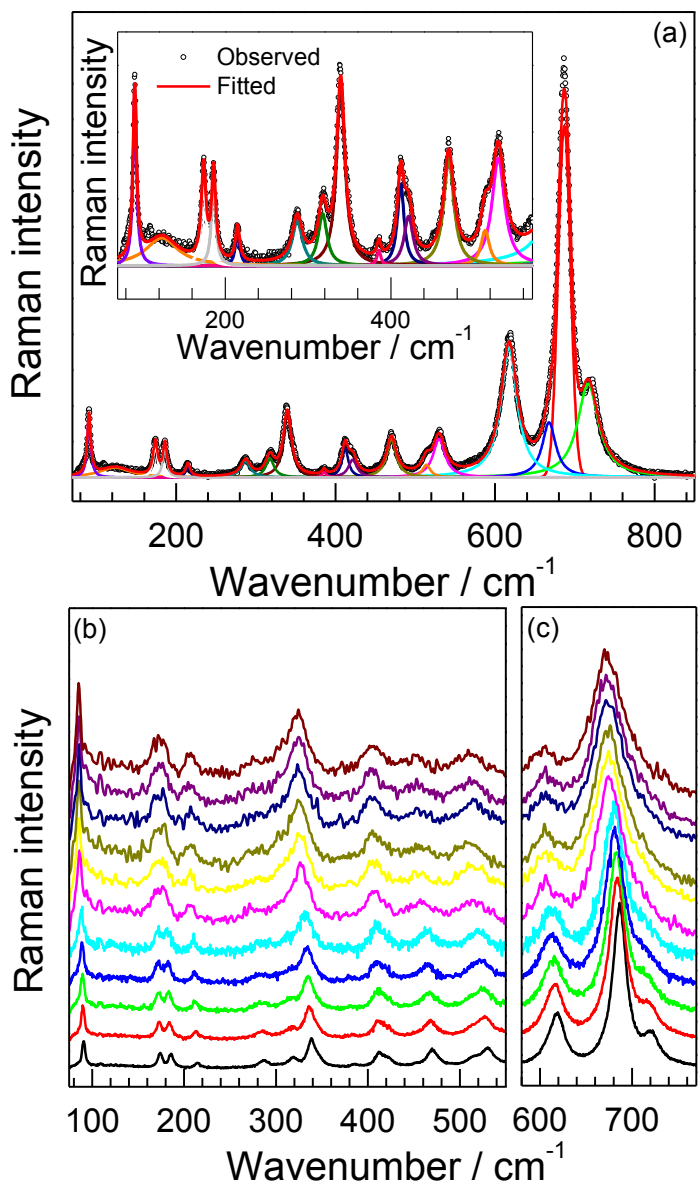

**Figure 1** – (a) Room-temperature Raman spectrum observed in BaM. (b-c) Temperature-dependent Raman spectra of BaM at low wavenumber (b) and high wavenumber (c) regions of the spectrum starting from 300 K (bottom – black line) up to 780 K (top – wine line) in steps of 50 K.

Table I – Assignment of the main Raman-active modes observed in BaM.

| Wavenumber | | Symmetry | Assignment |
| --- | --- | --- | --- |
| This Work | Kreisel *et al* [19] | | |
| 174 | 173 | $E_{1g}$ | Whole spinel block |
| 186 | 184 | $E_{1g}$ | Whole spinel block |
| 339 | 335 | $A_{1g}$ | Octahedra (mixed) |
| 412 | 409 | $A_{1g}$ | $Fe^{(5)}O_6$ octahedra dominated |
| 470 | 467 | $A_{1g}$ | Octahedra $Fe^{(1)}O_6$ and $Fe^{(5)}O_6$ |
| 618 | 614 | $A_{1g}$ | Octahedra $Fe^{(4)}O_6$ |
| 686 | 684 | $A_{1g}$ | Bipyramid $Fe^{(2)}O_5$ |
| 721 | 719 | $A_{1g}$ | Tetrahedra $Fe^{(3)}O^4$ |

Figures 1b and 1c show the temperature-dependent Raman spectra of BaM from room temperature up to 780 K. As expected, the Raman spectra do not show any subtle change since BaM does not show a structural phase transition in this temperature range. However, it is important point out that the Raman spectra keep a good intensity even at high temperatures. Also, clearly the intensity ratio between the 339 cm$^{-1}$ and 680 cm$^{-1}$ bands decreases significantly when the temperature increases.

The temperature dependence of the positions of some selected bands is shown in Figure 2. In Figure 2a and 2b we can see clearly an anomaly in the position of the Raman bands observed in 618 cm$^{-1}$ and 720 cm$^{-1}$ at 721 K, which is the Curie temperature of BaM, when BaM transforms into ferrimagnetic. Considering just anharmonic contributions to the temperature dependence of the phonon positions, they are modeled by Balkanski model [21], which shows that the position of a phonon depends on the temperature as

$$\omega(T) = \omega_o - C\left[1 + \frac{2}{(e^{\hbar\omega_o/k_B T} - 1)}\right] \quad (1)$$

with $C$ and $\omega_o$ being fitting parameters. Therefore, this anomaly is not expected once BaM has no structural phase transition at this temperature. Thus, this anomaly can be associated with two phenomena: a phonon renormalization due to a spin-phonon coupling; or due to a magnetostriction effect. However, any anomaly was observed in the lattice parameters of BaM when it undergoes this magnetic transition [22], which eliminates the possibility magnetostriction effect possibility. Thus, this anomaly is due to a spin-phonon coupling at the magnetic transition in BaM. A similar effect it was recently observed at low temperatures in BaM as reported by Xiang-Bai Chen *et al* [16]. The effect was stronger in these modes and was not observed in other phonons, as it is shown in Figures 2c and 2d.

Clearly, the anomaly observed in this phonon position changes the slope of the curve at $T_C$, as it is indicated by the dashed lines in Figure 2a and 2b, which suggest the coupling is significant. The spin-phonon coupling contribution $\Delta\omega_{s-ph}$ to the temperature dependence of the phonon position is given by [23]

$$\Delta\omega_{s-ph} = \lambda \langle \vec{S}_i \cdot \vec{S}_j \rangle \quad (2)$$

where $\lambda$ is the spin-phonon coupling constant, which can be positive or negative and it is different for each phonon; while $\langle \vec{S}_i \cdot \vec{S}_j \rangle$ is the scalar spin correlation function, which is negative in ferrimagnetic ordering case. Therefore, this observation indicates that, although the spin fluctuation is strong at temperatures higher than 200 K [16], the coupling is strong enough to be observed in this transition.

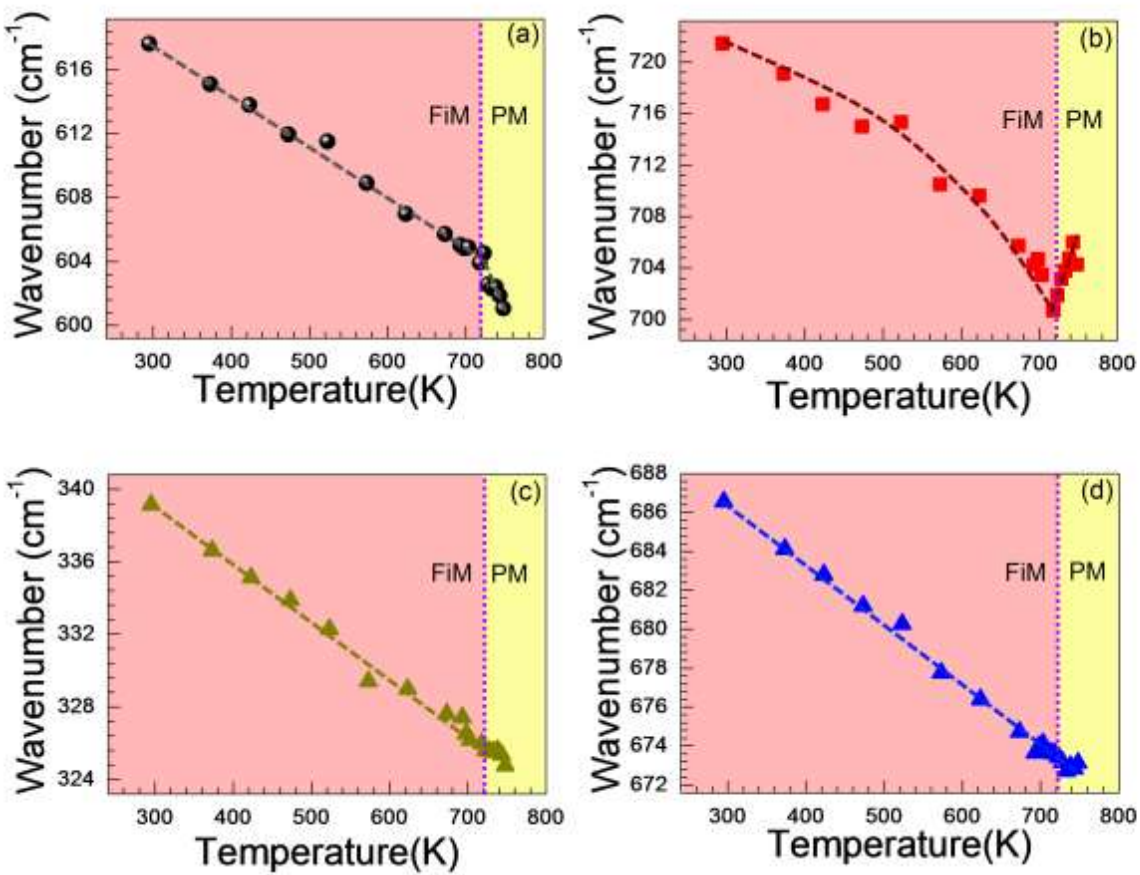

Figure 2 – Temperature dependence of the positions of some selected Raman-active modes. The dashed lines are guide for the eyes. The dotted line indicates the ferrimagnetic transition.

Also, this coupling observed in the 618 cm$^{-1}$ and 721 cm$^{-1}$ modes confirms the assumption raised by Xiang-Bai Chen *et al* [16], which proposed the spin-phonon coupling $\lambda$ is different for each phonon, even if they are coupled with same spin-spin interaction, since these authors do not observed a coupling in the phonon at 720 cm$^{-1}$,

9but on that in 618 cm$^{-1}$ and 680 cm$^{-1}$. Of course the behavior of each phonon, and consequently, of each coupling constant, depends on the relative motion of the phonon vibration in relation with the spin alignment.

In addition, an anomaly is observed in the FWHM of the mode observed at 680 cm$^{-1}$, as it is shown in Figure 3a. The temperature dependence of the FWHM of a Raman mode due to purely anharmonic effects is also given by Balkanski model as

$$\Gamma(T) = \Gamma_o \left[1 + \frac{2}{(e^{\hbar\omega_o/2k_B T}-1)}\right] \quad (3)$$

where $\Gamma_o$ is a fitting parameter. In the high temperature limit, this model predict a linear behavior for $\Gamma(T)$, as it is observed in Figure 3a until the ferromagnetic transition (see the dashed line which is a linear fitting of $\Gamma(T)$). Physically the FWHM is related to the phonon lifetime [24]. Thus, a slight change in the lattice due to the coupling changes slightly the phonons lifetime and, consequently, the FWHM. This effect is also frequently related to a spin-phonon coupling when observed at magnetic transitions [25] [26], confirming our assumption of a spin-phonon coupling in BaM at high temperatures.

Finally, as commented, there is a clear intensity reduction of the most intense phonon, observed at 680 cm$^{-1}$, in relation to the modes in the low wavenumber region. We plot the ratio between the intensities of the mode at 340 cm$^{-1}$ and 680 cm$^{-1}$, which is shown in Figure 3b. We see the ratio increases significantly with the temperature and, at the ferrimagnetic transition, it shows a clear anomaly.

In conclusion, detailed temperature-dependent Raman spectra of polycrystalline BaM were obtained at high temperatures from room temperature up to 780 K, which highlighted a spin-phonon coupling at the Curie temperature in the mode associated to the $Fe^{(4)}O_6$ octahedra and $Fe^{(3)}O_4$ tetrahedra. This observation confirmed the spin-phonon coupling is different for each phonon even when the phonons couple with the same spin configuration.

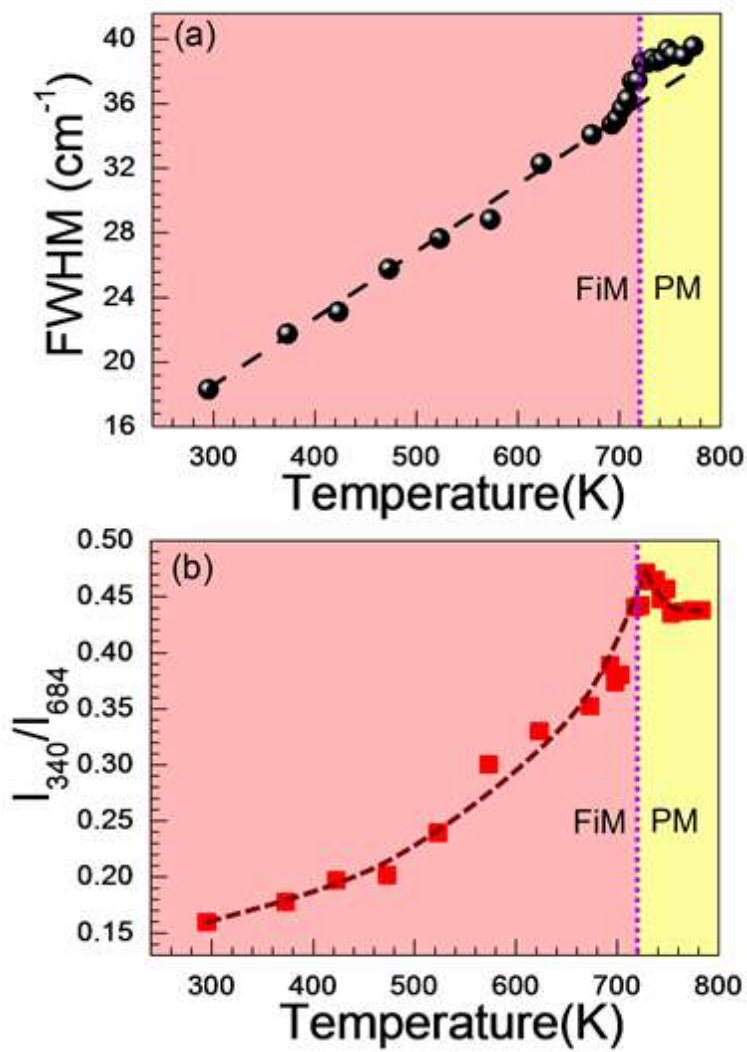

Figure 3 – (a) Temperature dependence of the FWHM of the mode observed at 680 cm$^{-1}$. (b) Ratio between the intensities of the modes observed at 340 cm$^{-1}$ and 680 cm$^{-1}$. The dashed lines are guide for the eyes. The dotted line indicates the ferromagnetic transition.

**Acknowledgments**

The authors acknowledge the financial support of the Brazilian funding agencies CAPES, CNPq and FAPEMA.

**References**

[1]   R. C. Pullar, Prog. Mater. Sci. **57**, 1191 (2012).

[2]   Ü. Özgür, Y. Alivov, and H. Morkoç, J. Mater. Sci. Mater. Electron. **20**, 789 (2009).

[3]   G. Turilli and A. Paoluzi, Ceram. Int. **19**, 353 (1993).

[4]   S. Capraro, J. P. Chatelon, M. Le Berre, H. Joisten, T. Rouiller, B. Bayard, D. Barbier, and J. J. Rousseau, J. Magn. Magn. Mater. **272-276**, E1805 (2004).

[5]   H. Pfeiffer, R. Chantrell, P. Görnert, W. Schüppel, E. Sinn, and M. Röslr, J. Magn. Magn. Mater. **125**, 373 (1993).

[6]   H. Sung, C.-J. Chen, W.-S. Ko, and H.-C. Lin, IEEE Trans. Magn. **30**, 4906 (1994).

[7]   H. Stäblein, in *Ferromagn. Mater. A Handb. Prop. Magn. Ordered, Vol. 3*, edited by E. Wohlfarth, 3rd ed. (North-Holland, 1982).

[8]   S. P. Marshall and J. B. Sokoloff, J. Appl. Phys. **67**, 2017 (1990).

[9]   G. Tan and X. Chen, J. Magn. Magn. Mater. **327**, 87 (2013).

[10]  Y. Tokunaga, Y. Kaneko, D. Okuyama, S. Ishiwata, T. Arima, S. Wakimoto, K. Kakurai, Y. Taguchi, and Y. Tokura, Phys. Rev. Lett. **105**, 257201 (2010).

[11]  R. B. Macedo Filho, A. Pedro Ayala, and C. William de Araujo Paschoal, Appl. Phys. Lett. **102**, 192902 (2013).

[12]  R. X. Silva, H. Reichlova, X. Marti, D. a. B. Barbosa, M. W. Lufaso, B. S. Araujo, a. P. Ayala, and C. W. a. Paschoal, J. Appl. Phys. **114**, 194102 (2013).

[13]  W. Ferreira, J. Agostinho Moreira, A. Almeida, M. Chaves, J. Araújo, J. Oliveira, J. Machado Da Silva, M. Sá, T. Mendonça, P. Simeão Carvalho, J. Kreisel, J. Ribeiro, L. Vieira, P. Tavares, and S. Mendonça, Phys. Rev. B **79**, 054303 (2009).

[14] K. Truong, M. Singh, S. Jandl, and P. Fournier, Phys. Rev. B **80**, 134424 (2009).

[15] M. N. Iliev, H. Guo, and A. Gupta, Appl. Phys. Lett. **90**, 151914 (2007).

[16] X.-B. Chen, N. T. Minh Hien, K. Han, J. Chul Sur, N. H. Sung, B. K. Cho, and I.-S. Yang, J. Appl. Phys. **114**, 013912 (2013).

[17] Y. P. Santos, B. C. Andrade, R. Machado, and M. a. Macêdo, J. Magn. Magn. Mater. **364**, 95 (2014).

[18] X. Zhang, Y. G. Zhao, Y. F. Cui, L. D. Ye, J. W. Wang, S. Zhang, H. Y. Zhang, and M. H. Zhu, Appl. Phys. Lett. **100**, 032901 (2012).

[19] J. Kreisel, G. Lucazeau, and H. Vincent, **137**, 127 (1998).

[20] J. Kreisel, S. Pignard, H. Vincent, J. P. Sénateur, and G. Lucazeau, Appl. Phys. Lett. **73**, 1194 (1998).

[21] M. Balkanski, R. Wallis, and E. Haro, Phys. Rev. B **28**, 1928 (1983).

[22] S. Chaudhury, S. K. Rakshit, S. C. Parida, Z. Singh, K. D. S. Mudher, and V. Venugopal, J. Alloys Compd. **455**, 25 (2008).

[23] E. Granado, A. García, J. Sanjurjo, C. Rettori, I. Torriani, F. Prado, R. Sánchez, A. Caneiro, and S. Oseroff, Phys. Rev. B **60**, 11879 (1999).

[24] L. Bergman, D. Alexson, P. Murphy, R. Nemanich, M. Dutta, M. Stroscio, C. Balkas, H. Shin, and R. Davis, Phys. Rev. B **59**, 12977 (1999).

[25] H. S. Nair, D. Swain, H. N., S. Adiga, C. Narayana, and S. Elzabeth, J. Appl. Phys. **110**, 123919 (2011).

[26] V. Srinu Bhadram, B. Rajeswaran, A. Sundaresan, and C. Narayana, EPL Europhysics Lett. **101**, 17008 (2013).